\documentclass[referee]{aa} % for a referee version
\usepackage{natbib}

\usepackage{graphicx}
\usepackage{txfonts}
\begin{document}
   \title{The inner Galactic bar traced by the VVV survey\thanks{Based on observations taken within the ESO VISTA Public Survey VVV, Programme ID 179.B-2002}}
   
   \author{O. A. Gonzalez$^{1}$  \and M. Rejkuba$^{1}$ \and D. Minniti$^{2,3,4}$ \and M. Zoccali$^{2}$ \and E. Valenti$^{1}$ \and R. K. Saito$^{2}$}
   
   \offprints{O. A. Gonzalez}
   \institute{ $^{1}$European
   Southern Observatory, Karl-Schwarzschild-Strasse 2, D-85748 Garching,
Germany\\ \email{ogonzale@eso.org; mrejkuba@eso.org; evalenti@eso.org}\\
   $^{2}$Departamento    Astronom\'ia    y Astrof\'isica,
   Pontificia Universidad  Cat\'olica de Chile,  Av. Vicu\~na Mackenna
   4860,         Stgo.,         Chile\\         \email{mzoccali@astro.puc.cl;
Dante@astro.puc.cl; rsaito@astro.puc.cl }\\
   $^{3}$Vatican Observatory, V00120 Vatican City State, Italy\\
   $^{4}$European Southern Observatory, Ave. Alonso de Cordova 3107, Vitacura, Santiago, Chile\\
}
   \date{Received / Accepted}

  
\abstract  
%  context (optional)  
{}
{We use the VVV survey observations in bulge regions close to the Galactic plane to trace the bar inclination at the Galactic latitude $b\sim\pm1$ and to investigate a distinct structure in the inner regions of the bar that was previously detected at positive latitude ($b=+1$).}
%  aims
{We use the $(J-K_s)$ colors of the red clump stars to obtain reddening values on $6\times6$ arcmin scale, minimizing the problems arising from differential extinction. Dereddened magnitudes are then used to build the luminosity function of the bulge in regions of $\sim$0.4 sq deg to obtain the mean red clump magnitudes. These are used as distance indicators to trace the bar structure.}
% method
{The luminosity function clearly shows the red clump mean magnitude variation with longitude, as expected from a large scale bar oriented towards us at positive Galactic longitude, with a dereddened magnitude varying from $K_{s_0}=13.4$ at $l=-10^\circ$ to $K_{s_0}=12.4$ at $l=+10^\circ$. We detect a change in the orientation of the bar in the central regions with $|l|<4^\circ$ at $b=\pm1^\circ$, in agreement with results obtained at positive latitudes by other authors. Our results are based on a different dataset and at different latitude, which shows that this change in the bar orientation is real. This suggests that there is an inner structure distinct to the large-scale Galactic bar, with a different orientation angle. This inner structure could be a secondary, inner bar, with a semi-major axis of $\sim500$ pc that is symmetric with respect to the Galactic plane.}
% results
{}
% conclusion
             
\authorrunning{Gonzalez et al.}
\titlerunning{Detection of the inner Galactic bar traced by the VVV survey}

\maketitle

%
%________________________________________________________________
\section{Introduction}

The presence of bars in spiral galaxies is common \citep[][]{eskridge00} and the Milky Way is no exception. Different studies based on various techniques have suggested that there is a large-scale bar with $\sim$2.5 kpc in radius oriented 15-30 degrees with respect to the Galactic center line of sight \citep[e.g.][]{dwek95,stanek97,bissantz02,babusiaux05,cabrera07,rattenbury07} with the near end towards positive longitudes. 

However, observations of external galaxies have led to the conclusion that the existence of double bars is not unusual \citep[e.g][]{laine02,erwin11} and thus have provided important constraints on their scales. Secondary bars are found to be generally small, with a typical size of $\sim500$ pc or a relative length of about 12\% of that of the main bar \citep[][]{erwin11}. Some observations in the very inner regions of the Milky Way have indeed suggested that there is another structure inside the Galactic bar, namely a nuclear bar \citep[e.g.][]{alard01,nishiyama05}. \citet[][]{alard01} used 2MASS star counts to show evidence of an excess in the inner regions of the Milky Way, which is interpreted as the presence of a small, lopsided, nuclear bar. In particular, \citet[][hereafter N05]{nishiyama05} used the magnitude of red clump (RC) stars at a latitude of $b=+1^\circ$ to measure the properties of the inner regions ($-10^\circ<l<10^\circ$) detecting a shallower inclination angle for the Galactic bar at fields with longitudes $|l|<4^\circ$. This was interpreted by N05 as the effect of a distinct inner bar structure. The existence of such a nuclear bar has strong effects on the gas distribution in the inner parts of the Galaxy. \citet[][]{rodriguez08} concluded that indeed a nuclear bar is likely to be responsible for the observed so-called central molecular zone. 

\begin{figure*}
\begin{center}
\includegraphics[scale=1.2,angle=0]{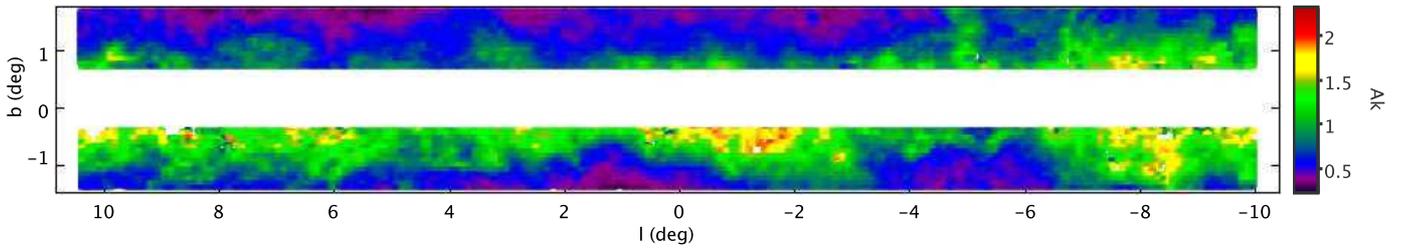}
\caption{K-band extinction map in Galactic coordinates showing the fields analyzed in this study. The 28 VVV tiles cover ranges of longitude $-10^\circ<l<10^\circ$ and of latitude $-1.4^\circ<b<-0.4^\circ$ and $0.7^\circ < b < 1.7^\circ$.}
\label{red}
\end{center}
\end{figure*}

In this Letter, we make use of the ESO Vista Variables in the Via Lactea survey  (VVV) to trace the magnitudes of RC giants in the longitude range $-10^\circ<l<10^\circ$ at $b=\pm1^\circ$ to constrain the morphology of the Galactic bar. The technique, based on using RC giants as accurate distance indicators across different lines of sight, was first applied by \citet[][]{stanek97} and adopted later by several authors to trace the bar morphology at different Galactic latitudes \citep[e.g.][]{babusiaux05, rattenbury07, cabrera07}. \citet{minniti11} applied this technique to the VVV survey data to detect the edge of the Galactic stellar disk. Here, we use it to test whether the change in the slope in the orientation of the Galactic bar observed by N05 at $b=+1^\circ$ is also evident in our data at $b=+1^\circ$, as well as at $b=-1^\circ$. We corroborate the measurements of N05, which suggests that there is a distinct structure in the inner regions of the Galactic bar that appears to be symmetric with respect to the Galactic plane.
%__________________________________________________________________

\section{The data}

For this study, we use the near-IR JK photometry from the VVV public survey\footnote{http://vvvsurvey.org}. A detailed description of the survey and the data can be found in \citet[][]{vvv10}. In particular, the photometric calibration and dereddened magnitudes were calculated as described in \citet[][]{gonzalez11b}, hence are only briefly described here. Multi-band catalogs were produced by a cross-match of sources between single-band catalogs produced at the Cambridge Astronomical Survey Unit (CASU) using the STILTS code \citep[][]{taylor06}. For this work, we analyzed a total of 28 tiles, containing about 30 million measured sources, which cover the survey region between $-1.4^\circ<b<-0.4^\circ$ and $+0.7^\circ<b<+1.7^\circ$ at longitudes across the range $-10^\circ<l<10^\circ$ (Fig.\ref{red}). The photometric calibration was obtained by cross-matching sources of VVV with those of 2MASS flagged with high quality photometry. This produced the final VVV J, H, and $K_s$ photometric catalogs fully consistent with the 2MASS photometric system. The calibrated VVV tile catalogues were then dereddened and used to build the luminosity functions for each of the tiles.

\section{Analysis and results}

The extinction correction was done following the prescriptions described in \citet[][]{gonzalez11b}. However, for the present analysis, we adopted the \citet[][]{nishiyama09} extinction law, where $A_k=0.528\cdot E(J-K_s)$ instead of the standard values of \citet[][]{cardelli89}, as the latter does not seem to be consistent with observations in the high-reddening regions analyzed here. Therefore, the dereddened $K_{s_0}$ magnitudes were obtained as
\begin{equation}
 K_{s_0}=K_s+0.528[(J-K_s)_0-(J-K_s)],
\end{equation} 
where we use the mean intrinsic RC color for the bulge of $(J-K_s)_0=0.68$ as measured in Baade's window \citep[][]{gonzalez11b}.
We note that by following this procedure we obtain reddening values in subfields of $6 \times 6$ arcmin instead of assigning individual extinction values to each star as in N05. Owing to the broadness of the bulge metallicity distribution \citep[e.g.][]{zoccali08,bensby11}, a single intrinsic RC color cannot be adopted for all stars but only to the mean of the observed RC color distribution in small subfields where the effects of differential extinction are minimized. Therefore, our procedure allows us to obtain dereddened $K_s$ magnitudes while keeping the imprints of the original bulge stellar population. Figure~\ref{red} shows the final extinction map and the field coverage for this study.

\begin{figure}
\begin{center}
\includegraphics[scale=0.82,angle=0]{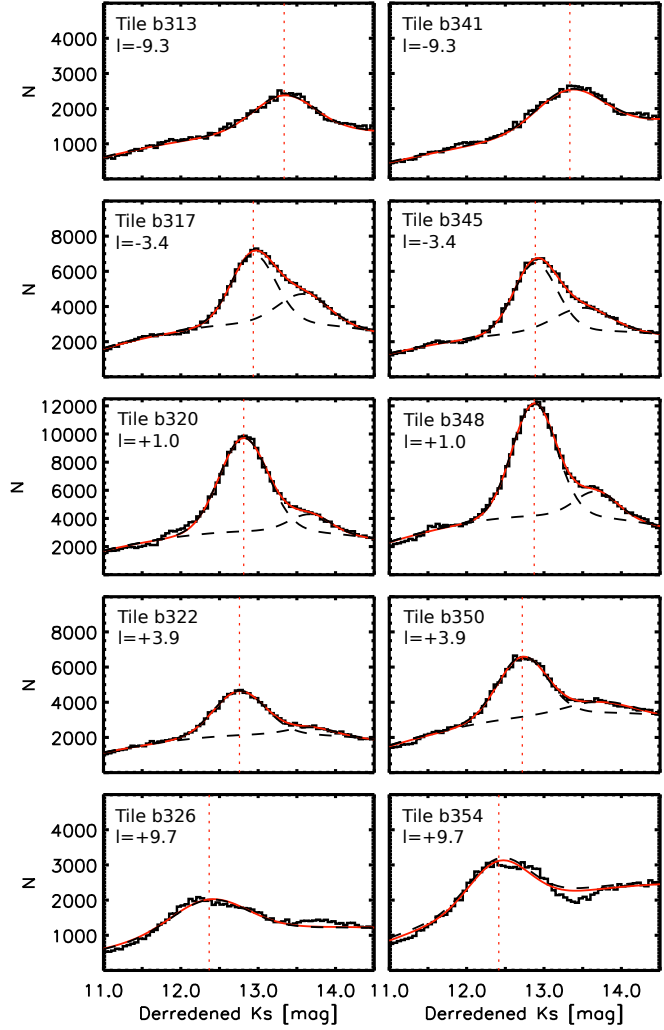}
\caption{$K_{s,0}$ luminosity functions for representative fields at $b=-1^\circ$ (left panels) and $b=+1^\circ$ (right panels). The red solid line shows the best fit to each distribution while the dashed line shows the individual Gaussian fits to the RC and to the additional peak when detected.}
\label{lfs}
\end{center}
\end{figure}

We applied a color cut $(J-K_s)_0>0.4$ to select the region of the color-magnitude diagram (CMD) dominated by bulge giants and avoid contamination from disk stars. Using the dereddened $K_{s_0}$ magnitudes of bulge giants, we then built the luminosity function for each tile and followed the method introduced by \citet[][]{stanek97} to measure the RC mean magnitudes in order to trace the mean distance of the bulge population. The base of the luminosity function was fitted with a second order polynomial and a Gaussian fit was then applied to measure the RC mean magnitude. Therefore the final fitting function is
\begin{equation}
N(K_{s_0})=a+bK_{s_0}+cK_{s_0}^2+\frac{N_{RC}} {\sigma_{RC}\sqrt{2\pi}}\exp\Big[-\frac{(K_{s_0}^{RC}-K_{s_0})^2}{2\sigma^2_{RC}}\Big].
\end{equation}
The results of the Gaussian fits for each RC are listed in Table~\ref{fits}. We also included an additional Gaussian fit for a peak in the luminosity function, distinct from the RC, which is centered at $K_{s_0}\sim13.7$. N05 mention the detection of this additional peak at a dereddened magnitude $K_{0}\sim$13.5 but they only performed Gaussian fits up to a magnitude where the RC is unaffected by this peak. \citet[][]{nataf11} presented its detection, based on optical OGLE data, and identified it as an anomalous red giant branch bump (RGBb). In \citet[][]{gonzalez11b}, the peak was observed using the VVV photometry up to $b=-4^\circ$. For larger distances from the plane, the detection of this peak becomes difficult because of the split in the RC \citep[][]{mcwilliam10}. However, we point out that, while the magnitude of the RC becomes brighter toward positive longitudes according to the bar orientation, the $K_{s_0}$ magnitude of this additional peak changes in the opposite direction. This is not expected if this feature corresponds to the RGBb. A possibility could be that it instead corresponds to RC stars from a more distant population. Assuming an intrinsic RC magnitude of $M_K= - 1.55$, the observed fainter RC peak would be at $\sim11.2$ kpc from the Sun. Further discussion of this point, such as the number counts of the RC with respect to those of the secondary peak cannot be addressed in this work because of the possible incompleteness at the magnitudes where this peak is observed, in particular for regions affected by high reddening and crowding. However, we assume here that the RC mean magnitudes remain reliable as long as the additional peak is included in the fitting procedure, independently of its origin. Figure~\ref{lfs} shows the results for the fits to the $K_{s,0}$ distribution of RC stars in representative fields of the regions studied in this work. 

Figure \ref{rc} shows the measured location of the Galactic bar obtained from the dereddened $K_{s_0}$ magnitude of the RC at $b=\pm1^\circ$, compared with those of N05 for $b=+1^\circ$ and adopting an intrinsic magnitude of the RC of $M_K=-1.55$ following \citet[][]{gonzalez11b}. Along every line of sight, the dispersion in distance was obtained from the measured $\sigma$ corrected by an intrinsic dispersion in the RC of 0.17 mag \citep{babusiaux05} and a photometric error of 0.05 mag. As discussed in \citet[][]{stanek94}, the observed orientation angle based on our method differs from the real one by an amount that depends on the thickness of the bar, measuring larger angles for a thicker bar. For this reason, we included in Figure~\ref{rc} examples of the observed orientations of a bar with axis ratios of $x/y=0.15$ \citep[][]{cabrera07} for the true angles of $15^\circ$, $30^\circ$, and $45^\circ$ following corrections shown in \citet[][]{stanek94}. Results are in good agreement with previous studies, which are consistent with the bar being located at 7.6 kpc, oriented at $30^\circ$, and with its near end towards positive longitudes. However, the position angle of the bar in the inner $\sim$1 kpc is more than $20^\circ$ larger than measured in the outer regions, which implies that there is an inner central structure.

\begin{table}
\caption{Fitting parameters for the RC and the additional peak} % title of Table
\label{fits} % is used to refer this table in the text
\centering % used for centering table
\setlength{\tabcolsep}{4pt}

{\footnotesize \begin{tabular}{c c c c c c c c} % centered columns (4 columns)
\hline\hline % inserts double horizontal lines
Tile & l & $K_{s_0}$ & peak & $\sigma$ & $K_{s_{0,2}}$ & peak$_2$ & $\sigma_2$\\ % table heading
     &[deg]&   [mag] &      &   [mag]  &   [mag]       &      &   [mag]       \\
\hline % inserts single horizontal line
$b=-1$\\
\hline % inserts single horizontal line
b313 & -9.3 & 13.36 & 1004.40 & 0.39 &   *    &   *     &  *  \\
b314 & -7.8 & 13.29 & 1219.88 & 0.42 &   *    &   *     &  *  \\
b315 & -6.3 & 13.25 & 2451.44 & 0.44 &   *    &   *     &  *  \\
b316 & -4.9 & 12.99 & 2569.64 & 0.30 &  13.53 & 1638.15 & 0.36\\ 
b317 & -3.4 & 12.94 & 3944.79 & 0.28 &  13.59 & 1683.01 & 0.32\\
b318 & -2.0 & 12.93 & 3894.66 & 0.29 &  13.68 & 930.18  & 0.22\\
b319 &  0.5 & 12.87 & 5241.86 & 0.30 &  13.69 & 821.97  & 0.21\\
b320 &  1.0 & 12.81 & 6658.11 & 0.31 &  13.68 & 1262.40 & 0.26\\
b321 &  2.4 & 12.80 & 4209.59 & 0.30 &  13.59 & 756.11  & 0.27\\
b322 &  3.9 & 12.76 & 2531.18 & 0.33 &  13.70 & 421.59  & 0.29\\
b323 &  5.3 & 12.69 & 1562.18 & 0.39 &  13.76 & 280.90  & 0.23\\
b324 &  6.8 & 12.57 & 978.848 & 0.39 &    *   &   *     &  *  \\
b325 &  8.3 & 12.42 & 793.638 & 0.40 &    *   &   *     &  *  \\
b326 &  9.7 & 12.37 & 980.697 & 0.48 &    *   &   *     &  *  \\
\hline % inserts single horizontal line
$b=+1$\\
\hline % inserts single horizontal line
b341 & -9.3 & 13.33 & 1142.66 & 0.43 &    *   &   *     &  *  \\
b342 & -7.8 & 13.23 & 1605.09 & 0.46 &    *   &   *     &  *  \\
b343 & -6.3 & 13.16 & 2013.17 & 0.49 &    *   &   *     &  *  \\
b344 & -4.9 & 13.10 & 2659.84 & 0.44 &    *   &   *     &  *  \\
b345 & -3.4 & 12.89 & 3930.44 & 0.28 &  13.51 & 1293.89 & 0.36\\
b346 & -2.0 & 12.92 & 7113.72 & 0.28 &  13.68 & 1817.71 & 0.29\\
b347 & -0.5 & 12.89 & 8410.80 & 0.27 &  13.66 & 1923.11 & 0.26\\
b348 &  1.0 & 12.87 & 8029.82 & 0.29 &  13.68 & 1829.86 & 0.27\\
b349 &  2.4 & 12.80 & 5536.06 & 0.30 &  13.66 & 1199.78 & 0.30\\
b350 &  3.9 & 12.72 & 3473.53 & 0.32 &  13.74 & 642.58  & 0.40\\
b351 &  5.3 & 12.71 & 3069.82 & 0.32 &    *   &    *    &  *  \\
b352 &  6.8 & 12.63 & 2062.99 & 0.36 &    *   &    *    &  *  \\
b353 &  8.3 & 12.53 & 1643.23 & 0.37 &    *   &    *    &  *  \\
b354 &  9.7 & 12.42 & 1321.51 & 0.39 &    *   &    *    &  *  \\
\hline %inserts single line
\end{tabular}
}
\tablefoot{Listed are the mean dereddened ($K_{s_0}$), sigma ($\sigma$) and peak of the Gaussian used to fit the RC in each one of the tiles. $K_{s_{0,2}}$, peak$_2$ and $\sigma_2$ are the parameters used fo fit the secondary peak.\\
}
\end{table}

\section{Discussion}

Recent studies seem to point out that secondary bars in barred spirals are indeed a common phenomenon, with at least 29\% of barred galaxies having a secondary bar \citep[e.g.][]{laine02}. Therefore, it would not be surprising that our Milky Way hosts such an inner structure. In this work, we have presented further evidence of the detection of an inner structure in the Galactic bar, as previously suggested by N05, identified by a change in the orientation angle of the bar measured at both positive ($b=+1$) and negative ($b=-1^\circ$) latitudes. since our results have been obtained independently and based on a different dataset from that of N05, we confirm that the detection is not an artifact owing to possible issues with either the photometry or calibrations. 

\begin{figure}
\begin{center}
\includegraphics[scale=0.58,angle=0]{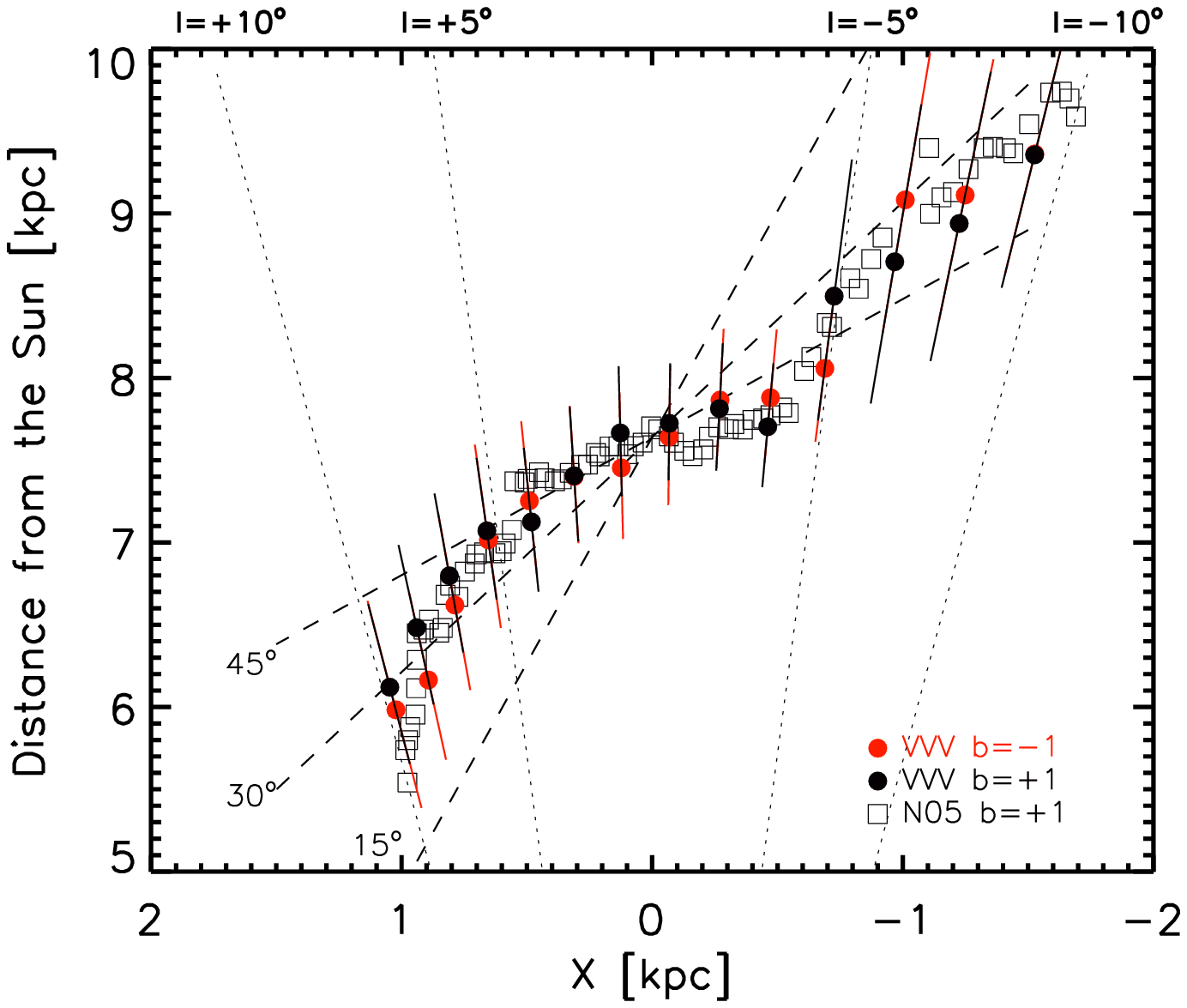}
\caption{Position of the Galactic bar with respect to the Sun as measured with the RC method assuming a mean RC magnitude of $M_K=-1.55$. Red and black filled circles show the results for VVV data at latitudes $b=-1^\circ$ and $b=+1^\circ$, respectively. Solid lines show the distance spread along each line of sight correcting for an intrinsic bulge dispersion of 0.17 mag and photometric errors. Black open squares are the results from N05 at $b=+1$. Dashed lines represent the observed orientation angles for true orientations of $15^\circ$, $30^\circ$, and $45^\circ$ following \citet[][]{stanek94}. Dotted lines show the lines of sight for longitudes $l=\pm5$ and $l\pm10$.}
\label{rc}
\end{center}
\end{figure}

However, we note that both studies are based on the use of the observed magnitude of the RC to trace distances across different regions of the bar. Therefore, one point to consider is the dependence of metallicity for the intrinsic magnitude of the RC \citep{girardi+salaris01}. This dependence can produce changes of $\sim$0.1 mag for 0.2 dex changes in [Fe/H]. However, these gradients in [Fe/H] have only been observed along the minor axis in the outer bulge \citep[][]{zoccali08} and are even more likely to be absent from the inner regions around $b\sim-1^\circ$ \citep[][]{rich_origlia07}. Even if gradients are indeed present also along the major axis for the inner regions considered here, to produce a similar effect to what we see in Fig.~\ref{rc}, the [Fe/H] gradient could not be \textit{radial}, but should go from one extreme of the bar to the other with a slope that changes in the inner longitudes $|l|<4^\circ$. These particular properties for the stellar populations of the Milky Way bar are very unlikely.

A second point to consider is the adopted extinction law for the determination of dereddened $K_{s_0}$ magnitudes. For a consistent comparison with N05 measurements, we have used the same extinction law based on \citet[][]{nishiyama09}. However, we note that this extinction law, which differs from the more commonly adopted laws  \citep[e.g.][]{cardelli89,savage79,rieke85}, was derived for the highly reddened inner bulge fields ($|b|<2^\circ$), while this does not seem to be the case for the whole bulge. In the study of RR Lyrae in the bulge from \citet[][]{kunder08}, variations with respect to the standard reddening law were observed in a few directions. However, they conclude that, on average, the extinction law of $R_V\sim3.1$ is consistent with the observations of the bulge at larger distances from the Galactic plane. Given this uncertainty in the correct extinction law with line of sight, a variation of it from standard values to that of \citet[][]{nishiyama09} could produce a change of up to 0.1 magnitudes in $K_{s_0}$ for the regions with the highest reddening ($A_k\sim 2.5$), as seen in Fig.~\ref{red}. However, these variations are likely to be randomly distributed along different lines of sight. That the same change in the bar orientation is now observed at both positive and negative latitudes, where reddening patterns differ greatly (Fig.~\ref{red}), leads us to conclude that our results are not an effect of extinction.

At larger Galactic latitudes ($|b|>3^\circ$), the flattening of the bar within $|l|<4^\circ$ is not observed. This, combined with the symmetric change in slope, leads us to suggest that we are detecting an inner second bar in the Milky Way. We cannot however exclude that the observed flattening is due to a particular change in the density distribution of RC stars in the inner bulge. 

%Finally, \citet[][]{lopez07} suggested that the observations of N05 may be an effect of the superposition of a long bar and triaxial bulge with different orientations \citep[e.g][]{hammersley00,benjamin05,lopez07,cabrera08}. We cannot discard this posibility with the present data, in particular because the long bar studies are done at larger longitudes than those analyzed here (l). However, is not clear how these two structures would coexist and how this would affect the inner regions \citep[][]{romero11}. Furthermore, as concluded by \citet[][]{rodriguez08}, the observed CMZ properties indeed support the presence of a inner bar.

\section{Conclusions}

We have used the VVV data at $b=\pm1^\circ$ to build the bulge luminosity functions along different lines of sight, and to measure the mean dereddened $K_{s_0}$ magnitude, which can be used as a distance indicator. The RC is clearly detected in all fields, and within $|l|<5^\circ$ a second component is detected in the K-band luminosity function. Given that the average magnitude shift of this second peak does not follow the primary RC, it could be an independent structure detected at $\sim11.2$ kpc from the Sun e.g. a spiral arm behind the bulge. 

The main RC traces the mean orientation for the Galactic bar, which produces a magnitude variation from $K_{s_0}\sim $13.4 ($\sim9.6$ kpc) at $l=-10$ to $K_{s_0}\sim $12.4 ($\sim6.2$ kpc) at $l=+10^\circ$. However, a different position angle is observed for the central $\sim$1 kpc regions with longitudes $-4^\circ<l<4^\circ$. This is in excellent agreement with the results of N05, which show the same change in the orientation of the bar at positive latitudes $b=+1^\circ$. These results provide additional evidence of a possible secondary inner bar with a semi-major axis of $\sim500$ pc, that is symmetric to the Galactic plane and has an orientation angle that differs from the large scale bar. Our measurements are consistent with an angle of the inner bar larger than $45^\circ$, although in these regions we measure the superposition of both the inner and the large scale bar and therefore detailed models are required to constrain the geometry and structure in the inner bulge. %Further analysis of stellar kinematics is required in order to obtain constrains to fully characterize the properties of this inner structure.

\begin{acknowledgements}
We thank O. Gerhard, I. Martinez-Valpuesta, and R. Benjamin for helpful discussions. We gratefully acknowledge the use of data products from the Cambridge Astronomical Survey Unit, and funding from
the FONDAP Center for Astrophysics 15010003, the BASAL CATA Center for
Astrophysics and Associated Technologies PFB-06, the MILENIO Milky Way
Millennium Nucleus from the Ministry of Economycs ICM grant P07-021-F and Proyectos FONDECYT Regular 1087258, 1110393 and 1090213. MZ is also partially supported by Proyecto Anillo ACT-86. RS acknowledges financial support from
CONICYT through GEMINI Project Nr. 32080016. We warmly thank the ESO Paranal Observatory staff for performing the observations and Mike Irwin, Eduardo Gonzalez-Solares, and Jim Lewis at CASU for pipeline data processing support. This material is based upon work supported in part by the National Science Foundation under Grant No. 1066293 and the hospitality of the Aspen Center for Physics. We thank the anonymous referee for comments that helped to improve the paper.
%%%%%%%%%%%%%%%%%%%%%%%%%%%%%%%%%%%%% 
%%%%%%%%%%%%%%%%%%%%%%%%%%%%%%%%%%%%%
\end{acknowledgements}

\renewcommand*{\bibfont}{\tiny}
\bibliographystyle{aa}

\bibliography{mybiblio}

\end{document}